\documentclass{kapproc} 
\usepackage{graphicx}
\usepackage{ae}
\usepackage{txfonts}
\usepackage{natbib}
\bibpunct{(}{)}{;}{a}{}{,}

\RequirePackage{graphicx}%
\RequirePackage{epsf}%

\upperandlowercase
\let\footnote\savefootnote
\let\footnotetext\savefootnotetext 
\let\footnoterule\savefootnoterule 
\newcommand{\sil}{\:\lower0.6ex\hbox{$\stackrel{\textstyle <}{\sim}$}\:}

\def\araa{{\em ARAA}}
\def\apj{{\em ApJ}}
\def\apjl{{\em ApJ}}
\def\apjs{{\em ApJS}}
\def\aap{{\em A\&A}}

\def\mnras{{\em MNRAS}}

\def\pasp{{\em PASP}}

\begin{document}


\articletitle{The stellar mass spectrum from non-isothermal gravoturbulent
  fragmentation}


\chaptitlerunninghead{IMF from non-isothermal fragmentation}


\author{Ralf Klessen\altaffilmark{1}, Katharina
  Jappsen\altaffilmark{1}, Richard Larson\altaffilmark{2}, Yuexing
  Li\altaffilmark{3,4}, Mordecai-Mark Mac Low\altaffilmark{3,4}}

\affil{\altaffilmark{1} Astrophysikalisches Institut Potsdam, An der Sternwarte 16, 14482
  Potsdam, Germany, \\ 
\altaffilmark{2}Department of Astronomy, Yale University, New Haven, CT
  06520-8101, U.S.A , \\
\altaffilmark{3} Department of Astrophysics, American
  Museum of Natural History, 79th Street at Central Park West, New
  York, NY 10024-5192, U.S.A.\\
\altaffilmark{4} Department of Astronomy,
  Columbia University, New York, NY 10027, U.S.A }

\email{rklessen@aip.de, akjappsen@aip.de, larson@astro.yale.edu,
  yuexing@amnh.org, mordecai@amnh.org}

 \begin{abstract}
   Identifying the processes that determine the initial mass function
   of stars (IMF) is a fundamental problem in star formation theory.
   One of the major uncertainties is the exact chemical state of the
   star forming gas and its influence on the dynamical evolution.
   Most simulations of star forming clusters use an isothermal
   equation of state (EOS).  We address these issues and study the
   effect of a piecewise polytropic EOS on the formation of stellar
   clusters in turbulent, self-gravitating molecular clouds using
   three-dimensional, smoothed particle hydrodynamics simulations. In
   these simulations stars form via a process we call gravoturbulent
   fragmentation, i.e., gravitational fragmentation of turbulent gas.
   To approximate the results of published predictions of the thermal
   behavior of collapsing clouds, we increase the polytropic exponent
   $\gamma$ from 0.7 to 1.1 at some chosen density $n_{\rm c}$, which
   we vary from from $4.3\times10^4\,\mathrm{cm^{-3}}$ to
   $4.3\times10^7\,\mathrm{cm^{-3}}$.  The change of thermodynamic
   state at $n_{\rm c}$ selects a characteristic mass scale for
   fragmentation $M_{\rm ch}$, which we relate to the peak of the
   observed IMF.  We find a relation $M_{\mathrm{ch}} \propto
   n_{\mathrm{c}}^{-0.5\pm0.1}$.  Our investigation supports the idea
   that the distribution of stellar masses largely depends on the
   thermodynamic state of the star-forming gas.  The thermodynamic
   state of interstellar gas is a result of the balance between
   heating and cooling processes, which in turn are determined by
   fundamental atomic and molecular physics and by chemical
   abundances. Given the abundances, the derivation of a
   characteristic stellar mass may thus be based on universal
   quantities and constants.
 \end{abstract}

\section{Introduction}

Although the IMF has been derived from vastly different regions, from
the solar vicinity to dense clusters of newly formed stars, the basic
features seem to be strikingly universal to all determinations
\citep{KRO01b}. Initial conditions in star forming regions can vary
considerably. If the IMF depends on the initial conditions, there
would thus be no reason for it to be universal.  Therefore a
derivation of the characteristic stellar mass that is based on
fundamental atomic and molecular physics would be more consistent.

There are many ways to approach the formation of stars and star
clusters from a theoretical point of view. In particular models that
connect stellar birth to the turbulent motions ubiquiteously observed
in Galactic molecular clouds have become increasingly popular in
recent years. See, e.g., the reviews by \citet{LAR03} and
\citet{MAC04}.  The interplay between turbulent motion and
self-gravity of the cloud leads to a process we call gravoturbulent
fragmentation: Supersonic turbulence generates strong density
fluctuations with gravity taking over in the densest and most massive
regions \citep[e.g.][]{LAR81, FLE82, PAD95, PAD97, KLE98, KLE00b,
  KLE01c, PAD02}. Once gas clumps become gravitationally unstable,
collapse sets in. The central density increases and soon individual or
whole clusters of protostellar objects form and grow in mass via
accretion from their infalling envelopes.

However, most current results are based on models that do not treat
thermal physics in detail. Typically, a simple isothermal equation of
state (EOS) is used. The true nature of the EOS, thus, remains a
major theoretical problem in understanding the fragmentation
properties of molecular clouds.
      
Recently \citet{LI03} conducted a systematic study of the effects of a
varying polytropic exponent~$\gamma$ on gravoturbulent fragmentation.
Their results showed that $\gamma$ determines how strongly
self-gravitating gas fragments. They found that the degree of
fragmentation decreases with increasing polytropic exponent $\gamma$
in the range $0.2 < \gamma < 1.4$ although the total amount of mass in
collapsed cores appears to remain roughly consistent through this
range. These findings suggest that the IMF might be quite sensitive to
the thermal physics. However in their computations, $\gamma$ was left
strictly constant in each case. Here we study the effects of using a
piecewise polytropic equation of state and investigate if a change in
$\gamma$ determines the characteristic mass of the gas clump spectrum
and thus, possibly, the turn-over mass of the IMF.

\section{Thermal Properties of Star-Forming Clouds}
\label{sec:therm-prop}

Observational evidence predicts that dense prestellar cloud cores show
rough balance between gravity and thermal pressure \citep{BEN89,
  MYE91}.  Thus, the thermodynamical properties of the gas play an
important role in determining how dense star-forming regions in
molecular clouds collapse and fragment.  Observational and theoretical
studies of the thermal properties of collapsing clouds both indicate
that at densities below $10^{-18}\,\mathrm{g\,cm^{-3}}$, roughly
corresponding to a number density of
$n=2.5\times10^5\,\mathrm{cm^{-3}}$, the temperature decreases with
increasing density. This is due to the strong dependence of molecular
cooling rates on density \citep{KOY00}. Therefore, the polytropic
exponent $\gamma$ is below unity in this density regime. At densities
above $10^{-18}\,\mathrm{g\,cm^{-3}}$, the gas becomes thermally
coupled to the dust grains, which then control the temperature by
far-infrared thermal emission. The balance between compressional
heating and thermal cooling by dust causes the temperature to increase
again slowly with increasing density. Thus the temperature-density
relation can be approximated with $\gamma$ above unity in this regime
\cite[Larson this volume; see also ][]{LAR85,SPA00}. Changing $\gamma$
from a value below unity to a value above unity results in a minimum
temperature at the critical density. As shown by \citet{LI03}, gas
fragments efficiently for $\gamma < 1.0$ and less efficiently for
higher $\gamma$. Thus, the Jeans mass at the critical density defines
a characteristic mass for fragmentation, which may be related to the
peak of the IMF.

\section{Numerical Approach}
To gain insight into how molecular cloud fragmentation the
characteristic stellar mass may depend on the critical density we
perform a series of smoothed particle hydrodynamics calculations of
the gravitational fragmentation of supersonically turbulent molecular
clouds using the parallel code GADGET designed by \citet{SPR01}. SPH
is a Lagrangian method, where the fluid is represented by an ensemble
of particles, and flow quantities are obtained by averaging over an
appropriate subset of SPH particles, see \citet{BEN90} and
\citet{MON92}.  The method is able to resolve large density contrasts
as particles are free to move, and so naturally the particle
concentration increases in high-density regions. We use the
\citet{BAT97} criterion the resolution limit of our calculations. It
is adequate for the problem considered here, where we follow the
evolution of highly nonlinear density fluctuations created by
supersonic turbulence. We replace the central high-density regions of
collapsing gas cores by sink particles \citep{BAT95}. These particles
have the ability to accrete gas from their surroundings while keeping
track of mass and momentum. This enables us to follow the dynamical
evolution of the system over many local free-fall timescales.

We compute models where the polytropic exponent changes from
$\gamma=0.7$ to $\gamma=1.1$ for critical densities in the range
$4.3\times10^4\,\mathrm{cm^{-3}} \le n_{\mathrm{c}} \le
4.3\times10^7\,\mathrm{cm^{-3}}$. Each simulation starts with a
uniform density distribution, and turbulence is driven on large
scales, with wave numbers $k$ in the range $1 \le k < 2$. We use the
same driving field in all four models. The global free-fall timescale
is $\tau_{\mathrm{ff}}\approx10^5\,\mathrm{yr}$. For further details see
\citet{JKLLM04}.

\section{Dependency of the Characteristic Mass }

\begin{figure*}[t!]
  \centering \includegraphics[width=10.0cm]{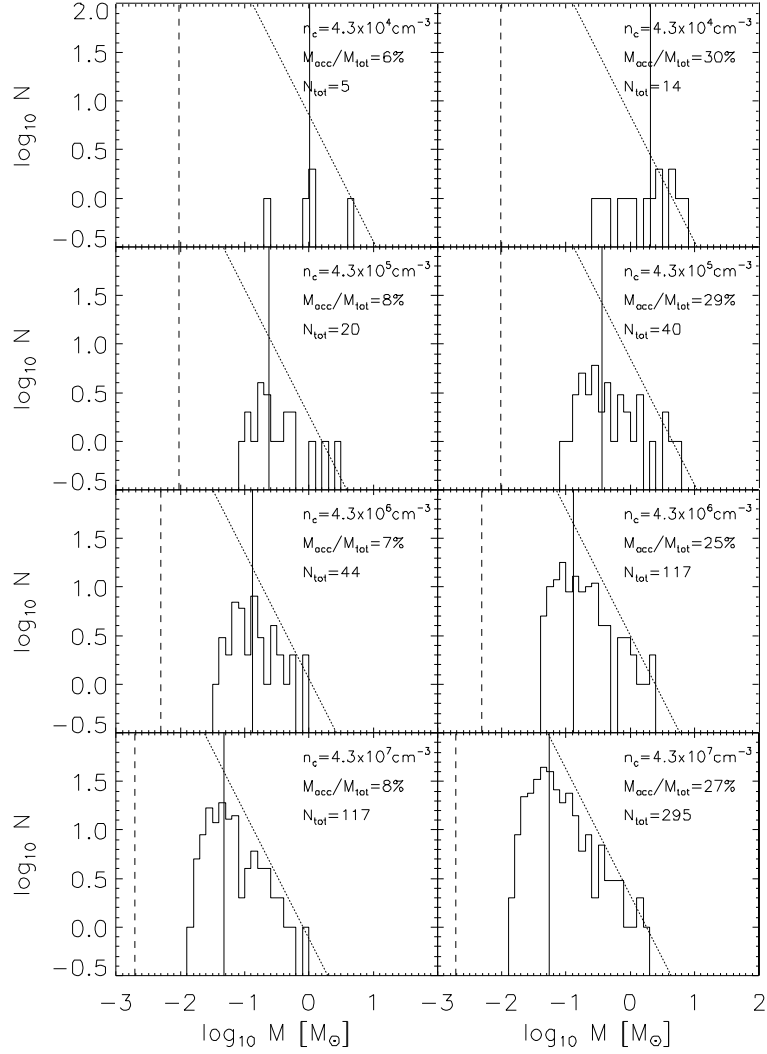}
\caption{Mass spectra of protostellar cores for four models with
  critical densities in the range $4.3\times10^4\,\mathrm{cm^{-3}} \le
  n_{\mathrm{c}} \le 4.3\times10^7\,\mathrm{cm^{-3}}$. We show two
  phases of evolution, when about 10\% and $30$\% of the mass has been
  accreted onto protostars. The {\em vertical solid line} shows the
  median mass of the distribution. The {\em dotted line} serves as a
  reference to the \citet{SAL55} slope of the observed IMF at high
  masses. The {\em dashed line} indicates our mass resolution limit.}
\label{fig:spectra}
\end{figure*}

To illustrate the effects of varying the critical density, we plot in
Fig.\ \ref{fig:spectra} the resulting mass spectra at different times
when the fraction of of mass accumulated in protostellar objects has
reached approximately 10\% and 30\%. This range of efficiencies
corresponds roughly to the one expected for regions of clustered star
formation \citep{LL03}. In the top-row model, the change in $\gamma$
occurs below the initial mean density. It shows a flat distribution
with only few, but massive cores. These reach masses up to
$10\,M_{\odot}$ and the minimum mass is about $0.3\,M_{\odot}$.  The
mass spectrum becomes more peaked for higher $n_{\mathrm{c}}$ and
shifts to lower masses.

We find closest correspondence with the observed IMF \citep{SCA98,
  KRO02, CHA03} for a critical density $n_{\rm c}$ of
$4.3\times10^6\,\mathrm{cm^{-3}}$ and for stages of accretion around
$30\%$. For high masses, the distribution exhibits a \citep{SAL55}
power-law behavior. For masses about the median mass the distribution
has a small plateau and then falls off towards smaller masses.

\begin{figure}[t!]
\centering  \resizebox{0.8\hsize}{!}{\includegraphics{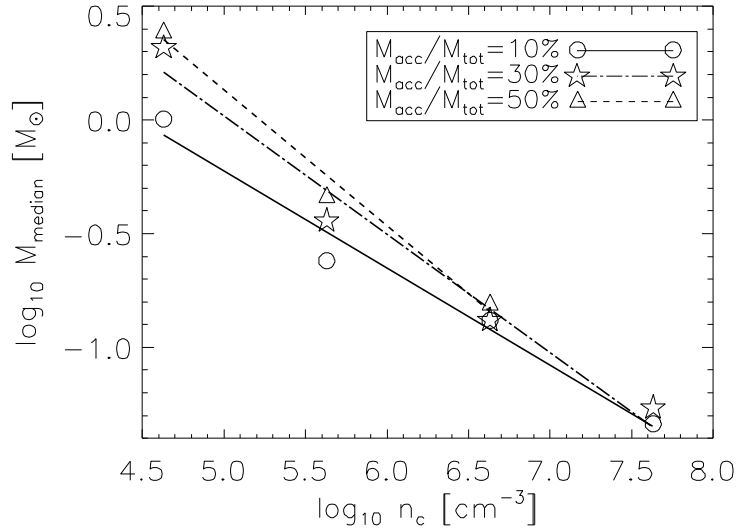}}
\caption{Plot of the  median mass of the protostellar cores $M_{\rm
    median}$ versus critical density $n_{\rm c}$.  We display results
  for different ratios of accreted gas mass to total gas
  mass~$M_{\mathrm{acc}}/M_{\mathrm{tot}}$, and   fit the data with
  straight lines. Their slopes take the values $-0.43\pm0.05$ ({\em
    solid line}), $-0.52\pm0.06$ ({\em dashed-dotted line}), and
  $-0.60\pm0.07$ ({\em dashed line}), respectively.}
\label{fig:all-med-rho}
\end{figure}

The change of median mass $M_{\rm median}$ with critical
density~$n_{\mathrm{c}}$ is quantified in Fig.~\ref{fig:all-med-rho}.
As $n_{\mathrm{c}}$ increases  $M_{\rm median}$ decreases. We fit our data
with straight lines. The slopes take values between $-0.4$ and $-0.6$.

\section{Discussion and Summary}
Using SPH simulations we investigate the influence of a piecewise
polytropic EOS on the gravoturbulent fragmentation of molecular
clouds. We study the case where the polytropic index $\gamma$ changes
from 0.7 to 1.1 at a critical density $n_{\mathrm{c}}$, and consider
the range $4.3\times10^4\,\mathrm{cm^{-3}} \le n_{\mathrm{c}} \le
4.3\times10^7\,\mathrm{cm^{-3}}$.

A simple scaling argument based on the Jeans mass $M_\mathrm{J}$ at
the critical density $n_\mathrm{c}$ leads to $M_{\mathrm{J}}\propto
n_{\mathrm{c}}^{-0.95}$ \citep[see][]{JKLLM04}. If there is a close
relation between the average Jeans mass and the gravoturbulent
fragmentation spectrum, a similar relation should hold for the
characteristic mass $M_{\mathrm{ch}}$ of protostellar cores.  Our
simulations qualitatively support this hypothesis, however, with the
weaker density dependency $M_{\mathrm{ch}} \propto
n_{\mathrm{c}}^{-0.5\pm0.1}$.  So indeed, the density at which
$\gamma$ changes from below unity to above unity defines a preferred
mass scale. Consequently, the peak of the resulting mass spectrum
decreases with increasing critical density. The distribution not only
shows a pronounced maximum but also a power-law tail towards higher
masses, similar to the observed IMF.

Altogether, supersonic turbulence in self-gravitating molecular gas
generates a complex network of interacting filaments. The overall
density distribution is highly inhomogeneous. Turbulent compression
sweeps up gas in some parts of the cloud, but other regions become
rarefied.  The fragmentation behavior of the cloud and its ability to
form stars depend on the EOS.  However, once collapse sets in, the
final mass of a fragment depends not only on the local Jeans
criterion, but also on additional processes. For example, protostars
grow in mass by accretion from their surrounding material.  In
turbulent clouds the properties of the gas reservoir are continuously
changing. And in addition protostars may interact with each other,
leading to ejection or mass exchange. These dynamical factors modify
the resulting mass spectrum, and may explain why the characteristic
stellar mass depends on the EOS more weakly than expected from simple
Jeans-mass scaling arguments.

Our investigation supports the idea that the distribution of stellar
masses depends, at least in part, on the thermodynamic state of the
star-forming gas.  If there is a low-density regime in molecular
clouds where the temperature $T$ sinks with increasing density $\rho$,
followed by a higher-density phase where $T$ increases with $\rho$,
fragmentation seems likely to be favored at the transition density
where the temperature reaches a minimum. This defines a characteristic
mass scale. The thermodynamic state of interstellar gas is a result of
the balance between heating and cooling processes, which in turn are
determined by fundamental atomic and molecular physics and by chemical
abundances. The theoretical derivation of a characteristic stellar mass may thus
be based on quantities and constants that depend mostly on the
chemical abundances in the star-lpwh klforming  cloud.


\end{document}